\documentclass[twocolumn,showpacs,preprintnumbers,amsmath,amssymb]{revtex4}

\usepackage{times}
\usepackage{natbib}
\usepackage{amssymb,amsbsy,amsmath,amsfonts}
\usepackage{graphicx}
\usepackage{float}
\usepackage{rotating}

\begin{document}
\title {Leading SU(3)-breaking corrections to the baryon magnetic moments \\
in Chiral Perturbation Theory }

\author{L. S. Geng$^1$}
\author{J. Martin Camalich$^1$}
\author{L. Alvarez-Ruso$^{1,2}$}
\author{M. J. Vicente Vacas$^1$}

\affiliation{$^1$Departamento de F\'{\i}sica Te\'orica and IFIC, Centro
Mixto, Institutos de Investigaci\'on de Paterna - Universidad de
Valencia-CSIC\\
$^2$Departamento de F\'isica, Universidad de Murcia, E-30071 Murcia, Spain}

\begin{abstract}
We calculate the baryon magnetic moments using covariant Chiral Perturbation Theory ($\chi$PT) within the Extended-on-mass-shell (EOMS) renormalization scheme. By fitting the two available low-energy constants (LECs), we improve the Coleman-Glashow description of the data when we include the leading SU(3) breaking effects coming from the lowest-order loops. This success is in dramatic contrast with previous attempts at the same order using Heavy Baryon (HB) $\chi$PT and covariant Infrared (IR) $\chi$PT. We also analyze the source of this improvement with particular attention on the comparison between the covariant results.
\end{abstract}
\date{\today}
\maketitle

In the limit that SU(3) is an exact flavour symmetry it is possible to relate the magnetic moments of the baryon-octet and the $\Lambda\Sigma^0$ transition to those of the proton and the neutron. These are the celebrated Coleman-Glashow formulas~\cite{Coleman:1961jn}. The improvement of this description requires the inclusion of a realistic SU(3)-breaking mechanism. Chiral Perturbation Theory ($\chi$PT), as the realization of non-perturbative QCD at low-energies~\cite{Gasser:1984gg,Gasser:1987rb,Scherer:2002tk}, should be an appropriate framework to tackle this problem in a systematic fashion. However, it was soon noticed that the leading-order chiral corrections are large and tend to worsen the results~\cite{Caldi:1974ta,Jenkins:1992pi}. This problem has often been used to question the validity of SU(3)-$\chi$PT in the baryon sector.

In the last decade several calculations in HB$\chi $PT  up to next-to-next-to-leading order (NNLO) have been performed both with~\cite{Jenkins:1992pi,Durand:1997ya,Puglia:1999th} and without~\cite{Meissner:1997hn} the inclusion of the baryon decuplet. The large number of LECs appearing at this order reduces the predictive power of the theory. Besides that, it is also known that there are substantial relativistic corrections~\cite{Krause:1990xc}. 

The development of covariant $\chi$PT has been troubled by the complications in the power counting introduced by the baryon mass as a new large scale~\cite{Gasser:1987rb}. Different ways of solving this problem, such as the IR~\cite{Becher:1999he} and, more recently, the  EOMS~\cite{Fuchs:2003qc} renormalization schemes, have been explored.
In SU(3) B$\chi$PT only the self-energies have been studied with both schemes~\cite{Ellis:1999jt,Lehnhart:2004vi}.
The baryon-octet magnetic moments have been calculated using the IR method~\cite{Kubis:2000aa} and, at NLO, the SU(3)-breaking corrections are still large. Moreover, the agreement with the data is even worse than in HB$\chi $PT. The size of NLO terms raises the question about the convergence of the chiral series~\cite{Durand:1997ya,Meissner:1997hn,Donoghue:2004vk}. 

In this letter we present a covariant calculation of the baryon-octet magnetic moments at $\mathcal{O}(p^3)$ (NLO) using the EOMS renormalization technique. In contrast to the previous works, we find small loop-corrections leading to a considerable improvement over the SU(3)-symmetric description. We also show the results in HB and covariant IR $\chi$PT, and investigate the origin of the differences. 

\begin{table*}
\caption{Coefficients of the tree-level [Eq. (\ref{eq:treeL})] and loop contributions [Eq. (\ref{eq:thirdO})] to the magnetic moments of the octet baryons.\label{table1}}
\begin{ruledtabular}
\begin{tabular}{cccccccccc}
 & $p$ & $n$ & $\Lambda$ & $\Sigma^-$ & $\Sigma^+$ & $\Sigma^0$ & $\Xi^-$ & $\Xi^0$ & $\Lambda\Sigma^0$  \\ 
\hline\hline 
$\alpha_B$ & $\frac{1}{3}$ & $-\frac{2}{3}$ & $-\frac{1}{3}$ & $\frac{1}{3}$ & $\frac{1}{3}$ & $\frac{1}{3}$  & $\frac{1}{3}$ & $-\frac{2}{3}$ & $\frac{1}{\sqrt{3}}$ \\

$\beta_B$ &  1 & 0 &  0 &  -1 &  1 &  0 &  -1 &  0 &  0  \\

$\xi^{(b)}_{B\pi}$ &  $  -(D+F)^2 $ & $(D+F)^2$ &   0 & $\frac{2}{3}(D^2+3F^2)$ & $-\frac{2}{3}(D^2+3F^2)$ &  0 & $(D-F)^2$ & $-(D-F)^2$ & $-\frac{4}{\sqrt{3}} DF$ \\

$\xi^{(b)}_{B K}$ & $-\frac{2}{3}(D^2+3F^2)$ & $-(D-F)^2$ & $ 2 DF$ & $ (D-F)^2$ & $ -(D+F)^2$ & $ -2 DF$ & $\frac{2}{3}(D^2+3F^2)$ & $ (D+F)^2$ & $-\frac{2}{\sqrt{3}} DF$\\

$\xi^{(c)}_{B\pi}$ & $ -\frac{1}{2}(D+F)^2$ & $ -(D+F)^2$ &  0  & $2 F^2$ & $ -2 F^2$ &  0 & $ \frac{1}{2}(D-F)^2$ & $ (D-F)^2$ & $\frac{4}{\sqrt{3}} DF$ \\

$\xi^{(c)}_{B K}$ & $ -(D-F)^2$ & $ (D-F)^2$ & $ -2 DF$ & $ (D+F)^2$ & $ -(D-F)^2$ & $ 2 DF$ &  $ (D+F)^2$ &  $ -(D+F)^2$ &  $\frac{2}{\sqrt{3}} DF$\\

$\xi^{(c)}_{B \eta}$ & $-\frac{1}{6}(D-3F)^2$ &  0 &  0 & $\frac{2}{3}D^2$ & $-\frac{2}{3}D^2$ &  0 & $\frac{1}{6}(D+3F)^2$ &  0 &   0 \\

\end{tabular}
\end{ruledtabular}
\end{table*}

In $\chi$PT, the power counting (PC) provides a systematic organization of amplitudes as a perturbative expansion in powers of $(p/\Lambda_{\chi\mathrm{SB}})^{n_{\chi PT}}$, where $p$ is a small momentum or scale and $\Lambda_{\chi\mathrm{SB}}$, the chiral symmetry breaking scale.
In the one-baryon sector, the chiral order of a properly renormalized diagram with $L$ loops, $N_M$($N_B$) meson (baryon) propagators, and $V_k$ vertices from $k$th-order Lagrangians, is $n_{\chi PT}=4L-2N_M-N_B+\sum_k k V_k$.
In the covariant theory with the $\overline{MS}$ renormalization prescription this rule is violated by lower-order analytical pieces~\cite{Ellis:1997kc,Fuchs:2003qc}. 

Different renormalization methods leading to a consistent PC have been developed within dimensional~\cite{Becher:1999he,Fuchs:2003qc} and cut-off~\cite{Djukanovic:2004px} regularization schemes. In the following we focus on the former ones.
In particular, the IR scheme~\cite{Becher:1999he} keeps the so-called infrared part of the loop function, which fulfills
 the PC and contains the non-analytic structures of the full function. The remaining so-called regular part, can be expanded close to the chiral limit in a series of analytic terms including the PC breaking pieces. They are then absorbed into the LECs of the most general (and infinite) Chiral Lagrangian. However, the IR formulation is known to introduce unphysical cuts at large momentum or meson masses~\cite{Becher:1999he,Bernard:2007zu}. On the other hand, in the EOMS scheme~\cite{Fuchs:2003qc} one subtracts from the full relativistic function just the PC breaking terms, absorbing them into a finite set of available lower-order LECs.

Our calculation requires the use of the standard lowest-order Chiral Lagrangians $\mathcal{L}^{(2)}_\phi$ and $\mathcal{L}^{(1)}_{\phi B}$, describing the pseudo-Goldstone bosons and baryons coupled to an external electromagnetic source (e.g.~\cite{Scherer:2002tk}).   
At second order there are two terms in the Chiral Lagrangian that contribute to the magnetic moments of the octet baryons
\begin{equation}
\mathcal{L}_{\gamma B}^{(2)}=\frac{b_6^D}{8 M_B}\langle\bar{B}\sigma^{\mu\nu}\lbrace F^+_{\mu\nu},B\rbrace\rangle+\frac{b_6^F}{8 M_B}\langle\bar{B}\sigma^{\mu\nu}[F^+_{\mu\nu},B]\rangle, \label{eq:BaryonLag2}
\end{equation}
where, in our case, $F^+_{\mu\nu}=2 |e| Q F_{\mu\nu}$, and $F_{\mu\nu}=\partial_\mu A_\nu-\partial_\nu A_\mu$ is the electromagnetic strength tensor. The LECs  $b_6^D$ and $b_6^F$ encode information about short-distance physics and should be determined from experiment within a given renormalization scheme. 
We take the values $D=0.80$ and $F=0.46$ for the axial and vector meson-baryon couplings appearing in $\mathcal{L}^{(1)}_{\phi B}$ and use the physical masses of the pseudoscalar mesons $m_\pi\equiv m_{\pi^\pm}=139.57$ MeV, $m_K\equiv m_{K^\pm}=493.68$ MeV and $m_\eta=547.5$ MeV. For the baryon mass we take a value of  $M_B=940$ MeV, so that the magnetic moments come expressed directly in nuclear magnetons. A moderate variation of $M_B$ is investigated below. As for the meson-decay constant in the chiral limit $F_0$, we choose an average between the physical values $F_\pi$=92.4 MeV, $F_K$=1.22$F_\pi$ and $F_\eta$=1.3$F_\pi$. Namely, $F_0\equiv F_\phi$=1.17$F_\pi$. The leading SU(3)-breaking corrections to the mass of the baryons in the octet and to the meson-decay constants contributes to the magnetic moments at higher orders.

\begin{figure}[t]
\includegraphics[width=\columnwidth]{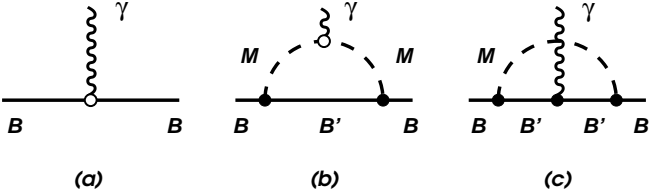}
\caption{Feynman diagrams contributing to the baryon anomalous magnetic moment. The solid lines correspond to baryons, dashed lines to mesons and the wiggly line denotes the external photon field. Black and white dots indicate $\mathcal{O}(p)$ and $\mathcal{O}(p^2)$ couplings respectively. \label{fig_diagram}}
\end{figure}

The Feynman diagrams for the anomalous magnetic moments up to $\mathcal{O}(p^3)$ are shown in Fig.~\ref{fig_diagram}. The tree-level coupling \textbf{\textit{(a)}} is given by the Lagrangian (\ref{eq:BaryonLag2}), and carries the leading-order (LO) result
\begin{equation}
\kappa_B^{(2)}=\alpha_B b_6^D+\beta_B b_6^F, \label{eq:treeL}
\end{equation}
where the coefficients $\alpha_B$ and $\beta_B$ for each of the baryons in the octet are listed in Table \ref{table1}. This lowest-order contribution is nothing else but the SU(3)-symmetric prediction leading to the Coleman-Glashow relations ~\cite{Coleman:1961jn, Jenkins:1992pi}.

The $\mathcal{O}(p^3)$ diagrams \textbf{\textit{(b)}} and \textbf{\textit{(c)}} account for the leading SU(3)-breaking corrections that are induced by the corresponding breaking in the masses of the pseudoscalar meson octet. Their contributions to the anomalous magnetic moment of a given member of the octet $B$ can be written as
\begin{eqnarray}
\kappa^{(3)}_B&=&\frac{1}{8\pi^2 F_\phi^2}\left(\sum_{M=\pi,K}\xi_{BM}^{(b)} H^{(b)}(m_M)\right.\nonumber\\
&+&\left.\sum_{M=\pi,K,\eta}\xi_{BM}^{(c)} H^{(c)}(m_M)\right)\label{eq:thirdO}
\end{eqnarray}
with the coefficients $\xi_{BM}^{(b,c)}$ listed in Table \ref{table1}. The loop-functions read
\begin{eqnarray}
 H^{(b)}(m)&=&-M_B^2+2 m^2+\frac{m^2}{M_B^2}(2 M_B^2-m^2)\log\left(\frac{m^2}{M_B^2}\right)\nonumber \\
&+&\frac{2m\left(m^4-4 m^2 M_B^2+2 M_B^4\right)}{M_B^2\sqrt{4M_B^2-m^2}}\,\arccos\left(\frac{m}{2\,M_B}\right), \nonumber \\
H^{(c)}(m)&=&M_B^2+2m^2+\frac{m^2}{M_B^2}(M_B^2-m^2)\log\left(\frac{m^2}{M_B^2}\right)\nonumber \\
&+&\frac{2m^3\left(m^2-3 M_B^2\right)}{M_B^2\sqrt{4M_B^2-m^2}}\,\arccos\left(\frac{m}{2\,M_B}\right). \label{eq:loop}
\end{eqnarray}
These loop integrals are convergent and do not depend on a renormalization scale. For the case of the proton and neutron this result coincides with the one obtained using a linearized form of the Gerasimov-Drell-Hearn sum rule~\cite{Holstein:2005db}. One also notices that they contain pieces $\sim M_B^2$ that contribute at $\mathcal{O}(p^2)$ to the magnetic moments, breaking the PC. 
\begin{table*}
\caption{Numerical results of the fits of $\tilde{b}_6^D$ and $\tilde{b}_6^F$ to the experimental values of baryon-octet magnetic moments up to $\mathcal{O}(p^3)$ in different $\chi$PT approaches. The experimental values with the corresponding errors are also displayed in the last row. All the values for the magnetic moments are expressed in units of nuclear magnetons, while $\tilde{b}_6^D$ and $\tilde{b}_6^F$ are dimensionless. \label{table2}}
\begin{ruledtabular}
\begin{tabular}{ccccccccccccc}
 & $p$ & $n$ & $\Lambda$ & $\Sigma^-$ & $\Sigma^+$ & $\Sigma^0$ & $\Xi^-$ & $\Xi^0$ & $\Lambda\Sigma^0$ & $\tilde{b}_6^D$ & $\tilde{b}_6^F$ & $\tilde{\chi}^2$ \\ 
\hline
\multicolumn{13}{c}{$\mathcal{O}(p^2)$}  \\
\hline
Tree level & 2.56 & -1.60 & -0.80 & -0.97 & 2.56 & 0.80 & -1.60 & -0.97 & 1.38 & 2.40 & 0.77 & 0.46  \\
\hline
\multicolumn{13}{c}{$\mathcal{O}(p^3)$}  \\
\hline
HB  & 3.01 & -2.62 & -0.42 & -1.35 & 2.18 & 0.42 & -0.70 & -0.52 & 1.68 & 4.71 & 2.48 & 1.01  \\

IR  & 2.08 & -2.74 & -0.64 & -1.13 & 2.41 & 0.64 & -1.17 & -1.45 & 1.89 & 4.81 & 0.012 & 1.86 \\

EOMS & 2.58 & -2.10 & -0.66 & -1.10 & 2.43 & 0.66 & -0.95 & -1.27 & 1.58 & 3.82 & 1.20 & 0.18  \\
\hline
Exp. &  2.793(0) & -1.913(0) & -0.613(4) & -1.160(25) & 2.458(10) & --- & -0.651(3) &-1.250(14) & $\pm$  1.61(8) & \multicolumn{3}{c}{---}  \\
\end{tabular}
\end{ruledtabular}
\end{table*}

In order to get rid of the PC problems we follow the EOMS scheme, by which these pieces are absorbed into the available counter-terms, $b_6^D$ and $b_6^F$. This is equivalent to redefining these two LECs as 
\begin{equation*}
\tilde{b}_6^D=b_6^D+\frac{3 D F M_B^2}{2\pi^2F_\phi^2},\;\;\;\;\;\;\;\;  \tilde{b}_6^F=b_6^F,
\end{equation*}
so that
\begin{eqnarray}
 \tilde{H}^{(b)}=H^{(b)}+M_B^2,  \;\;\;\;\;\;\;\tilde{H}^{(c)}=H^{(c)}-M_B^2. \label{eq:EOMS}
\end{eqnarray}
In this way, we have obtained the leading one-loop relativistic contribution to the magnetic moments starting from $\mathcal{O}(p^3)$. Furthermore, one is able to recover the leading non-analytical quantum correction in the HB formalism by setting $M_B\sim\Lambda_{\chi\mathrm{SB}}$
\begin{eqnarray}
\tilde{H}^{(b)}(m)\simeq\pi m M_B+\mathcal{O}(p^2),\;\;\; \tilde{H}^{(c)}(m)\simeq\mathcal{O}(p^2). \label{eq:HeavyB}
\end{eqnarray}
When added to the tree-level terms, this result completes the $\mathcal{O}(p^3)$ estimation of the baryon magnetic moments in the HB$\chi$PT approach~\cite{Jenkins:1992pi, Meissner:1997hn}. 

The IR amplitudes have been calculated in Refs.~\cite{Kubis:2000aa}. They can be obtained subtracting from the full loop-functions (\ref{eq:loop}) the corresponding regular parts, which can be expressed around the chiral limit as
\begin{eqnarray}
R^{(b)}(m)&=&-M_B^2+\frac{19 m^4}{6 M_B^2}-\frac{2 m^6}{5 M_B^4}+\cdots\,\,\,, \nonumber\\
R^{(c)}(m)&=&M_B^2+2 m^2+\frac{5m^4}{2 M_B^2}-\frac{m^6}{2 M_B^4}+\cdots\,\,\,. \label{eq:regular}
\end{eqnarray}
On the other side, the regular parts have unphysical cuts at $m=2M_B$. In short, in order to recover the PC, the IR formulation alters the analytical structure of the full relativistic theory~\cite{Holstein:2005db} such that the applications of this scheme for large meson masses (physical or unphysical) may become questionable~\cite{Bernard:2007zu}.
Nevertheless, since the differences between the full relativistic and IR results (or
those obtained in any other consistent scheme) are analytical in quark mass, they
should be reconciled with the adjustment of higher-order counter-terms.

In Table \ref{table2} we show the numerical results for the baryon magnetic moments obtained by minimizing  $\tilde{\chi}^2=\sum (\mu_{th}-\mu_{exp})^2$ as a function of $\tilde{b}_6^D$ and $\tilde{b}_6^F$. The $\Lambda\Sigma^0$ transition moment is not included in the fit but it is a prediction to be confronted with the experimental value. Moreover, we compare the tree-level result with the  $\mathcal{O}(p^3)$ loop results given by the three different $\chi$PT approaches discussed above, namely, the semi-relativistic HB Eq. (\ref{eq:HeavyB}) and the covariant Eq. (\ref{eq:loop}), within the EOMS Eq. (\ref{eq:EOMS}) or the IR Eq. (\ref{eq:regular}) renormalization schemes \footnote{The differences between the results we show for the HB and Covariant IR approaches and those in Ref. \cite{Kubis:2000aa} come from the different value for the baryon mass considered there ($M_B$=1151 MeV), as well as from the different treatment of the meson-decay constants. Namely, additional higher-order SU(3)-breaking corrections are introduced in \cite{Kubis:2000aa} by assigning physical values to $F_M$ ($M=\pi, K, \eta$).}. The experimental values of the magnetic moments are also displayed for comparison. 

The HB results show the longstanding problem of the poor convergence of $\chi$PT for the baryon magnetic moments. The leading non-analytical correction to the SU(3)-symmetric prediction amounts up to 80\% of the leading contribution for some of the baryons. In this approach, it is necessary to come up to $\mathcal{O}(p^4)$ to achieve reasonable convergence, although the role of the loop contributions is not clear in a scenario where one has the same number of parameters as of experimental values to fit~\cite{Durand:1997ya}. One expects that the covariant theory, with the proper higher-order chiral terms, should overcome the problem of convergence. However, the IR results are even worse that those obtained in HB. In particular, the quantum correction to the $\Sigma^-$ magnetic moment is three times bigger than the leading order one. The inclusion of NNLO is then required to achieve a successful description \cite{Kubis:2000aa}.

 The EOMS results presented in this work show an unprecedented NLO improvement over the tree-level description within dimensionally regularized $\chi$PT. Indeed, the $\tilde{\chi}^2$ in this approach is much better than those obtained with HB and IR. Moreover, it is also better than the tree level SU(3)-symmetric description . The convergence of the chiral expansion in our case can be accessed by separating the $\mathcal{O}(p^2)$ from the $\mathcal{O}(p^3)$ contributions for each magnetic moment (in units of nuclear magnetons)
\begin{eqnarray}
\mu_p=3.47\,\left( 1-0.257\right),\;\; \mu_n=-2.55\, \left( 1-0.175\right), \nonumber\\
\mu_\Lambda=-1.27\, \left(1-0.482\right),\;\; \mu_{\Sigma^-}=-0.93\,\left( 1+0.187 \right), \nonumber\\ \mu_{\Sigma^+}=3.47\,\left( 1-0.300 \right),\;\;\; \mu_{\Sigma^0}=1.27\, \left(1-0.482\right),\nonumber\\
\mu_{\Xi^-}=-0.93\,\left( 1+0.025 \right), \;\; \mu_{\Xi^0}=-2.55\,\left( 1-0.501 \right),\nonumber\\
\mu_{\Lambda\Sigma^0}=2.21\,\left( 1- 0.284\right).\;\;\;\;\;\;\;\;\;\;\;\;\;\;\;\;\;\;\nonumber
\end{eqnarray}
In dramatic contrast with the HB~\cite{Meissner:1997hn} and IR results~\cite{Kubis:2000aa}, we find that the NLO term represents, at most, a half of the leading contribution. This is consistent with the expected maximal correction of about $m_\eta/ \Lambda_{\chi\mathrm{SB}}$. Remarkably, we obtain a value for $\mu_{\Lambda\Sigma^0}$ very close to the experimental one assuming a positive sign.
\begin{figure}[t]
\includegraphics[scale=0.65]{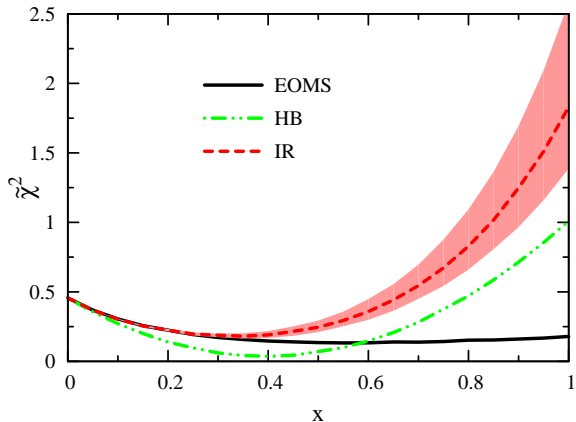}
\caption{(Color on-line) SU(3)-breaking evolution (see text for details) of the minimal $\tilde{\chi}^2$ in the $\mathcal{O}(p^3)$ $\chi$PT approaches under study. We also show the shaded areas produced by the uncertainty in $M_B$ when varying from 0.8 GeV to 1.1 GeV. This effect lies within the line thickness in the EOMS case, while the HB is insensitive to it.   \label{fig_graph}}
\end{figure}

In order to understand the differences between the three $\chi$PT formulations, we study the evolution of the minimal $\tilde{\chi}^2$ as we {\it switch-on} the SU(3)-breaking effects, by introducing the parameter $x=m_M/m_{M,phys}$ (where $M=\pi, K, \eta$) and varying it between zero and one. As seen in Fig.\ref{fig_graph}, the three approaches coincide in the vicinity of the chiral limit. The EOMS and IR results stay very close up to $x\sim0.4$. As $x$ increases further HB and IR description of data get worse while, on the contrary, the EOMS result lies well below the SU(3) symmetric one. Following the analysis of Ref.~\cite{Holstein:2005db}, we interpret the unrealistic IR behaviour as a manifestation of the change of the analytical structure of the theory made in this formulation. Certainly, this is due to the fact that in SU(3)-$\chi$PT one has to deal with $K$ and $\eta$ mesons which have masses larger than 350 MeV, the limit deemed acceptable for meson masses in one-loop SU(2)-$\chi$PT calculations~\cite{Bernard:2007zu}. 

We have also studied the uncertainties of our results to the particular value chosen for $M_B$. The shaded areas in the plot are produced by varying $M_B$ in the interval of $0.8$ GeV$\leq M_B\leq $1.1 GeV. While in HB the result is independent of the value of this parameter and IR manifests a clear sensitivity to it (the fit being worse for larger $m/M_B$ ratios), the EOMS result presents an intriguing insensitivity to $M_B$ (the shaded area lies within the thickness of the solid curve in Fig.\ref{fig_graph}). As pointed out in Ref.~\cite{Holstein:2005db}, this feature, as well as the soft dependence on the SU(3)-breaking exhibited by the EOMS curve, is due to subtle cancellations encoded into the full relativistic results.

In summary, we have improved the SU(3)-symmetric description of the baryon-octet magnetic moments by including the leading quantum effects provided by  relativistic $\chi$PT within the EOMS scheme. Besides the relativistic corrections, analyticity has proved to be of fundamental importance. Indeed, the effect of the unphysical cuts embedded into the IR loops containing  $K$- and $\eta$-mesons shows to be sizable. In addition to the first successful description of baryon-octet magnetic moments at NLO, this work contributes to clarify the longstanding puzzle regarding the applicability of baryon $\chi$PT in the SU(3)-flavor case. A careful study of different SU(3)-flavor observables is required in order to establish to what extent the improved convergence of EOMS with respect to IR found in this work is a general feature.

This work was partially supported by the  MEC grant  FIS2006-03438 and the EU Integrated Infrastructure
Initiative Hadron Physics Project contract RII3-CT-2004-506078. L.A.R. aknowledges financial support from the Seneca Foundation, L.S.G. from the MEC in the Program 
``Estancias de doctores y tecnologos extranjeros''. J.M.C. acknowledges the same institution for a FPU grant.

\end{document}